\documentclass[prb,aps,twocolumn,superscriptaddress,preprintnumbers,floats,amsmath,amssymb]{revtex4}

\usepackage{graphicx}
\usepackage{dcolumn}
\usepackage{bm}
\usepackage{amssymb}
\usepackage{psfrag}
\usepackage[latin1]{inputenc}

\def \be{\begin{equation}}
\def \ee{\end{equation}}

\def\a{\sigma}

\begin{document}

\title{The energy scale behind the metallic behaviors in low-density Si-MOSFETs}

\author{Genevi\`{e}ve Fleury}
\affiliation{SPEC-IRAMIS, CEA Saclay F-91191 Gif-sur-Yvette Cedex, France\\}
\author{Xavier Waintal}
\affiliation{SPEC-IRAMIS, CEA Saclay F-91191 Gif-sur-Yvette Cedex, France\\}
\affiliation{SPSMS-INAC, CEA 17 rue des Martyrs, 38054 Grenoble CEDEX 9, France}
\date{\today}

\begin{abstract} We show that the unexpected metallic behavior (the so-called two-dimensional metal-insulator transition) observed in low-density Silicon metal-oxide-semiconductor field-effect transistors (Si-MOSFETs) is controlled by a unique characteristic energy scale, the polarization energy. On one hand, we perform Quantum Monte Carlo calculations of the energy needed to polarize the 
two dimensional electron gas at zero temperature, taking into account Coulomb interactions,
valley degeneracy and electronic mobility (disorder). On the other hand, we identify the characteristic
energy scale controlling the physics in eight different sets of experiments. 
We find that our {\it ab-initio } polarization energies
(obtained without any adjustable parameters) are in perfect agreement with the observed characteristic energies for all
available data, both for the magnetic field and temperature dependence of the resistivities.
Our results put strong constraints on possible mechanisms responsible for the metallic behavior. In particular, there are strong 
indications that the system would eventually become insulating at low enough temperature.
\end{abstract}

\maketitle
\section{Introduction}
Until 1994, two dimensional systems were widely believed to be ultimately insulators~\cite{abrahams1979} as the presence of 
even a tiny disorder is enough to make the one-body electronic wavefunctions localized. In fact, until then, the experimental 
corpus was widely consistent with this paradigm\cite{bishop1980,uren1980}: the electrical resistance was found to increase as one 
lowered the temperature, indicating an eventual divergence at zero temperature. The observation of metallic behaviors in 
low-density Silicon metal-oxide-semiconductor field-effect transistors (Si-MOSFETs)\cite{kravchenko1994} hence came as a 
surprise and have been puzzling the community ever since~\cite{kravchenko2004,pudalov2004}. The experiments were first 
acknowledged with skepticism but they were soon repeated and extended by several groups\cite{kravchenko1995,kravchenko2000,mertes1999,popovic1997,pudalov1998}, including 
in other materials\cite{hanein1998bis,ribeiro1999,papadakis1998,coleridge1997,hanein1998,mills1999,simmons1998,lai2005}, so that it became clear that some intrinsic physics of the two-dimensional electron 
gas was probed in those experiments.

On the theory side, many different scenarios were proposed. In the weak disorder regime, perturbative calculations performed in the eighties\cite{altshuler1980,finkelshtein1983,castellani1984} showed the crucial role of spin and valley degeneracy and more recently, calculations in the limit of an infinite number of valleys suggested that electronic interactions could stabilize a non-Fermi liquid metallic phase\cite{punnoose2005}. Other theoretical works proposed that the metal-insulator transition could be due to an emulsion between a Wigner crystal and a Fermi liquid\cite{spivak2003} or to some sort of a Wigner-Mott transition\cite{camjayi2008}. It has also been argued that the metallic behaviors could be the signature of a superconducting phase\cite{belitz1998,phillips1998,thakur1998}. In contrast to these rather "extreme" scenarios other works proposed a more  conservative picture. This includes temperature dependent screening\cite{sarma1999,klapwijk1999}, temperature dependent disorder\cite{altshuler1999,altshuler2000ter} or temperature dependent scattering between two different spin sub-bands split by a strong spin-orbit coupling\cite{papadakis1999,murzin1998,yaish2000}. Percolation phenomenons were also claimed to play a role at the metal-insulator transition itself\cite{meir1999,shi1999,he1998,leturcq2003,sarma2005,manfra2007}. 
The problem was also approached using numerical techniques and a delocalization effect by interactions was observed repeatedly\cite{waintal1999,berkovitz2002,kotlyar2001,vojta1998,srinivasan2003,denteneer1999}.

The large spread in the above proposals, together with rather intense controversies between the conservative and 
the less conservative scenarios naturally lead to some confusion in the field so that no scientific consensus has 
emerged so far. 
The goal of this paper is twofold. First, we will perform a critical analysis of the existing experimental data and show 
that extremely strong constraints can be put on the possible theoretical scenarios. Second, following our recent proposal\cite{fleury2008bis}, we will show that a simple model with a minimum number of (well established) ingredients is enough to capture all the salient experimental facts.

In section~\ref{exp}, we summarize the main experimental facts with a focus on the characteristic energies that can be 
extracted from the data.
In section~\ref{model}, we introduce our minimum model to describe Si-MOSFETs, which is then used in section~\ref{qmc} 
to calculate the 
polarization energy $E_p$ in presence of both disorder and electron-electron interactions. In section~\ref{exp2}, 
we come back to the 
experimental data and show that our calculated $E_p$ agrees quantitatively (with no adjustable parameter) with 
the characteristic energies 
that can be extracted from either the magneto-resistance data or the resistance versus temperature data. Finally, 
we discuss the implications 
of the above findings for our own scenario\cite{fleury2008bis} (section~\ref{scenario}) as well as 
others (section~\ref{conclusion}).

\section{Basic experimental facts}\label{exp}

\subsection{Main observations}
\begin{figure}[!h]
\includegraphics[width=8.5cm]{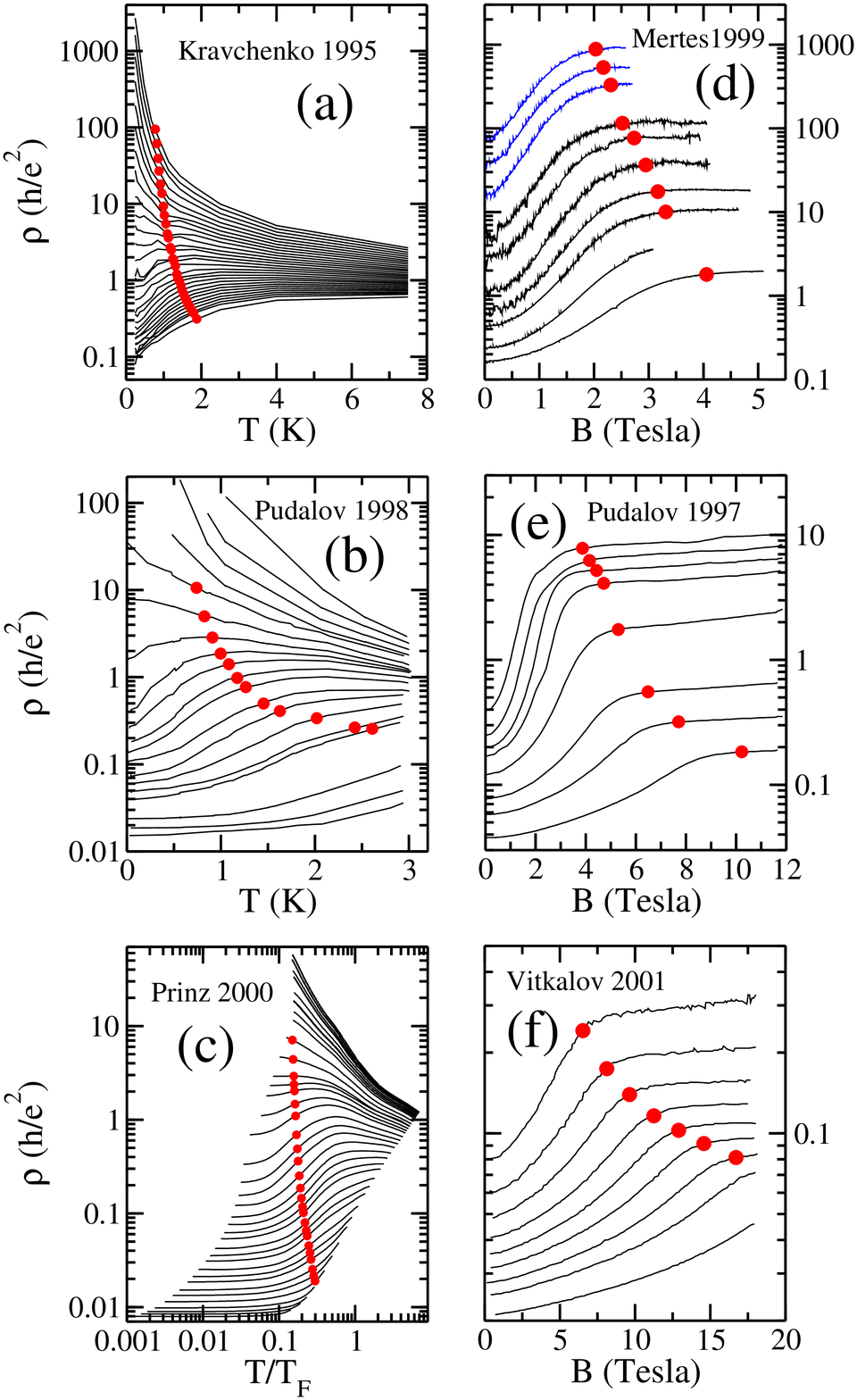}
\caption{\label{rho_exp} 
$\rho$ as a function of $T$ (left panels) and as a function of $B$ (right panels), for different $n_s$, in various Si-MOSFETs ((a) from Ref.~\cite{kravchenko1995}, (b) from Ref.~\cite{pudalov1998}, (c) from Ref.~\cite{prinz2000}, (d) from Ref.~\cite{mertes1999}, (e) from Ref.~\cite{pudalov1997} and (f) from Ref.~\cite{vitkalov2001}). The red circles are the corresponding polarization temperatures $T_p$ and polarization magnetic fields $B_p$ obtained from the numerical simulations.}
\vskip -0.4cm
\end{figure}

Let us quickly review the basic experimental facts that we want to understand. 
The first set of measurements that actually started the interest for these high mobility MOSFETs is the behavior of the
resistivity $\rho(T)$ as a function of temperature (at temperatures lower than a few Kelvin where phonon scattering no 
longer comes into play). The $\rho(T)$ behavior is well understood in two limiting cases. At high density, the resistivity is rather small and depends only weakly on temperature, except 
for a weak negative $\partial\rho/\partial T$ due to weak localization. 
At low density, the resistivity is much larger and one 
eventually reaches a clear insulating behavior where $\rho$ depends very strongly (exponentially) on temperature 
($\partial\rho/\partial T<0$). Before the report made in Ref.~\cite{kravchenko1994}, the common behavior observed in 
Si-MOSFETs, as well as in other two dimensional electron gases was a simple crossover between the high and the low 
density limit. Hence, the observation of an intermediate density regime
with $\partial\rho/\partial T>0$ (hereafter referred as property {\bf P1}) came as a large surprise. In the left panels of Fig.~\ref{rho_exp}, we reproduced the data of three different experiments, including the original set of data from Kravchenko \textit{et al}. One can observe a rather 
pronounced $\partial\rho/\partial T>0$ as 
the resistivity increases by a factor 10 between $100\,\textup{mK}$ and $4\,\textup{K}$. It was naturally surmised that in this regime, the 
resistivity could be extrapolated to a finite value at zero temperature, so that the system was in a metallic 
state, in contradiction with the prediction of scaling theory of localization\cite{abrahams1979}.

The second important raw experimental feature is the dependence of the resistivity versus an {\it in-plane} magnetic field $B$: in presence of a magnetic field, the resistivity increases quickly before saturating (property {\bf P2}). This is illustrated in the right panels of Fig.~\ref{rho_exp} for three different experiments. The increase of resistivity is comparable to the increase that one gets with increasing temperature, and $\partial\rho/\partial T$ quickly becomes negative (insulating like).

In our opinion, {\bf P1} and {\bf P2} are the two main experimental features that require an explanation. There have been
an important effort to establish scaling laws and critical exponents close to the metal-insulator transition, but we will not discuss this aspect. Indeed, once the metallic behavior {\bf P1} is understood, there is no question that, as the system must be eventually insulating at low density, some sort of transition must occur. While this transition might be of interest, it is not fundamentally puzzling in itself.

\subsection{First analysis}

One of the questions that many physicists had in mind when the first
experiments~\cite{kravchenko1994} came up was, what's new in those devices? 
Why have we not seen this behavior before? In fact, there was nothing
qualitatively new about those MOSFETs. 
The real novelty lied in the fact that the samples were of extremely high mobility (for Si-MOSFETs) which 
allowed the experiments to be carried out at much lower densities than previously possible. 
Working at lower densities has three consequences: 

{\bf Property P3}. The ratio $r_s$ of interaction over kinetic energy is large. 
This ratio $r_s=m^*e^2/(4\pi\epsilon_0\epsilon\hbar^2\sqrt{\pi n_s})$ 
(where $n_s$ is the electronic density, $m^*=0.19\,m_e$ the effective mass, $e$ the 
electron charge, $\epsilon_0$ the dielectric constant and $\epsilon=7.7$ the relative dielectric constant)
takes values between $5-10$ for the metallic region, hence the
 system is intrinsically correlated. 

\begin{figure}[!h]
\begin{center}
\psfrag{AAA}{\Large $1/\sqrt{k_Fl}$}
\psfrag{BBB}{\Large $r_s$}
\psfrag{Uren80}{\tiny\cite{uren1980}}
\psfrag{Si-14 de KRAV94}{\tiny Si-14 in \cite{kravchenko1994}}
\psfrag{(3) de VIT01}{\tiny (3) in \cite{vitkalov2001}}
\psfrag{MERTES99}{\tiny\cite{mertes1999}}
\psfrag{Si-12b de KRAV95}{\tiny Si-12b in \cite{kravchenko1995}}
\psfrag{Si-15 de PUD98}{\tiny Si-15 in \cite{pudalov1998}}
\psfrag{Si-12 de KRAV94}{\tiny Si-12 in \cite{kravchenko1994}}
\psfrag{Si-15 de KRAV94}{\tiny Si-15 in \cite{kravchenko1994}}
\psfrag{Bishop80}{\tiny\cite{bishop1980}}
\psfrag{Krav00}{\tiny\cite{kravchenko2000}}
\includegraphics[keepaspectratio, width=8.5cm]{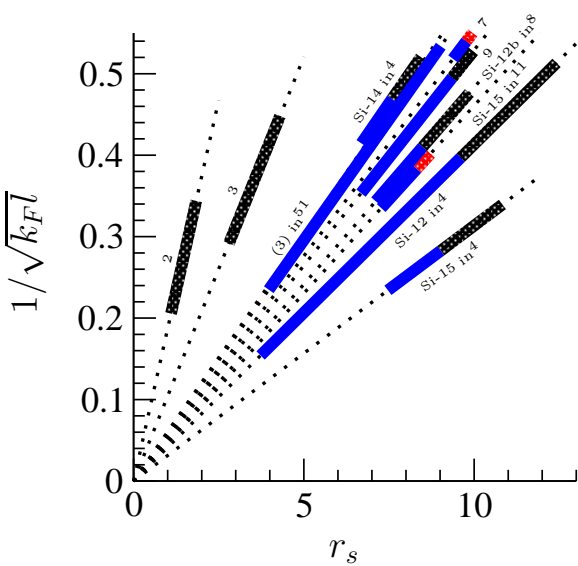} 
\caption{\label{fig_diagphaseexp}
Experimental $r_s$ versus $1/\sqrt{k_Fl}$ phase diagram of the Si-MOSFET, constructed from ten samples studied in the literature. Each sample corresponds to a straight line in the diagram (dotted lines). Blue (dashed red) refers to the density range where metallic $\partial\rho/\partial T>0$ (insulating $\partial\rho/\partial T<0$) behaviors are observed. This diagram should be compared with Fig.4 of Ref.\cite{fleury2008bis}.
         } 
\end{center}
\vskip -0.4cm
\end{figure}

{\bf Property P4}. The effective disorder seen by the electrons, as parametrized
by $1/k_Fl$ ($k_F$ Fermi momentum and $l$ mean free path) is paradoxically not smaller: indeed the
high mobility of the samples is compensated by the fact that the density is lower and therefore $k_Fl\propto n_s$ remains of the order of a few units. Note that contrary to $n_s$ (hence $r_s$) which can be almost directly measured with Shubnikov-de-Haas 
measurements, there is no generic way to measure the disorder.
At high density, $r_s\ll 1$ and the interaction effects can be neglected, so that $k_Fl$ is simply related to the conductance
$g$ of the system, $g= (2e^2/h) k_Fl$. In this limit, we estimate the product $\eta\equiv r_s\sqrt{k_F l}$ which should remain
constant as one lowers the density (as $r_s\propto 1/\sqrt{n_s}$ and $k_Fl\propto n_s$). Hence, $\eta$ is estimated at large density by the non-interacting formula $\eta=\sqrt{\mu} e^{3/2} m^* /(4\pi \hbar^{3/2} \epsilon_0\epsilon)$. For real samples where disorder have intrinsic characteristic lengths, the mobility $\mu$ may have a density dependence. Typical values of $\eta$ are of the order of $20$. In Fig.~\ref{fig_diagphaseexp}, we have collected the trajectories in the $r_s$ versus $1/\sqrt{k_Fl}$ phase diagram for ten different samples: as one lowers the density, the samples follow a straight line $1/\sqrt{k_Fl}=r_s/\eta$. The regions where the samples have a metallic behavior are plotted in blue while the insulating behaviors appear in dashed red. When comparing the samples where the metallic behavior is observed with the older samples that did not show such a behavior (the two on the left), one indeed observes that the metallic behavior is observed at larger $r_s$ ({\bf P3}) but in the same regime of disorder ({\bf P4}).

{\bf Property P5}. The Fermi energy $E_F$ decreases so that the ratio $T/T_F$ increases (where $E_F=kT_F=\pi\hbar^2n_s/2m^*$, $T_F$ is the Fermi temperature). Typical experimental values are quite large,
 $T/T_F\approx20\%$ around the inflexion point of the $\rho(T)$ curves as can be seen 
in Fig.~\ref{rho_exp}\,(c). This last point, raised in Ref.~\cite{altshuler2000bis}
 has been overlooked by many authors and is however crucial: the metallic behavior happens for
temperatures which are of the same order of magnitude as the Fermi temperature. Hence, the usual ``low
energy'' paradigm where one only looks at electrons close to the Fermi surface does not hold in those
devices.

It is actually the purpose of this paper to identify the characteristic energy that controls the behavior of
 $\rho(T)$ and $\rho(B)$.  As we shall see, both are controlled by a unique energy scale, 
the polarization energy $E_p$ of the Si-MOSFET. 

\section{A minimum model for Silicon MOSFETs}\label{model}
We now introduce the model that we will use for our calculations. It includes Coulomb repulsion, disorder, spin and a twofold valley degeneracy (present in Si-MOSFETs but not in their GaAs counterparts, Silicon being an indirect gap semi-conductor). This model is minimum in the sense that the first three ingredients are already known to be relevant experimentally, as discussed in the above section, while the fourth one is known to be present and we will show that it plays an important role for quantitative predictions.
 The system contains $N\gg 1$ particles in a lattice made
of $L_x\times L_y$ sites, the particles being
equally split up into the two degenerate valleys, with $N_{\uparrow}$ spin up and $N_{\downarrow}$ spin down. 
The spin configuration defines the system polarization 
\be
p=\frac{N_{\uparrow}-N_{\downarrow}}{N_{\uparrow}+N_{\downarrow}}.
\ee
The Hamiltonian for the $4$-components plasma reads,
\be
\label{eq:model}
H=-t\sum_{\langle\vec r\vec r'\rangle \a}c_{\vec r \a}^\dagger c_{\vec r' \a} + \sum_{\vec r \a}
v_{\vec r}n_{\vec r \a}
+\frac{U}{2} \sum_{\vec r \a\ne\vec r'\a'} V_{\vec r-\vec r'} n_{\vec r \a} n_{\vec r' \a'}
\ee
where $c_{\vec r \a}^\dagger$ and $c_{\vec r \a}$ are the usual creation and annihilation operators of 
one electron on site $\vec r$ with inner degree of freedom $\a$, 
the sum $\sum_{\langle\vec r\vec r'\rangle\a}$ is restricted to nearest 
neighbors and $n_{\vec r \a}=c_{\vec r \a}^\dagger c_{\vec r \a}$ is the density operator. The internal degree
of freedom $\a=1\dots 4$ corresponds to the spin and valley degeneracy degrees of freedom.
The disorder potential $v_{\vec r}$ is uniformly distributed inside $\left[-W/2,W/2\right]$. 
$t$ is the hopping parameter and $U$ is the effective strength of the two body interaction $V_{\vec r}$. To reduce finite size 
effects, $V_{\vec r}$ is obtained~\cite{waintal2006} from the bare Coulomb interaction using the Ewald summation 
technique,
\be
V_{\vec r} = \sum_{\vec L} \frac{1}{|\vec r +\vec L|} {\rm Erfc}(k_c |\vec r +\vec L|)
\ee
$$+\frac{2\pi}{L_x L_y} \sum_{\vec K\ne \vec 0} 
\frac{1}{|\vec K|} 
{\rm Erfc}(|\vec K|/(2 k_c))\cos (\vec K\cdot\vec r)\,.
$$
In the previous equation, $k_c$ is a (irrelevant) cut off. The vector $\vec L$
takes discrete values $\vec L= (n_x L_x,n_y L_y)$ with $n_x$ and $n_y$ integer numbers.
The vector $\vec K$ also takes discrete values, $\vec K= (\frac{2 \pi}{L_x} n_x,
\frac{2 \pi}{L_y} n_y)$ and $(n_x,n_y)\ne (0,0)$. The complementary error function is
defined as ${\rm Erfc}(x)=\frac{2}{\sqrt{\pi}} \int_x^\infty e^{-t^2} dt$.
We work at small filling 
factor $\nu\equiv N/(L_x L_y)\ll 1$, where we recover the continuum limit. The two dimensionless parameters that control the physics read,
\be
r_s=U/(2 t \sqrt{\pi \nu}), \ \ \ \ \ k_F l = 48\pi \nu t^2/W^2\,.
\ee

\section{Quantum Monte-Carlo calculation of the polarization energy.} \label{qmc}
 This section is devoted to the numerical calculation of the polarization energy of the $4$-components plasma, as a function of $r_s$ and $k_Fl$. We compute the ground state energy per particle $E(p)$ of the system with the Green function  Quantum Monte Carlo (QMC) technique in the Fixed Node approximation (see Ref.~\cite{waintal2006} for the numerical method). Particular care was given to the extrapolation to the thermodynamics limit ($N\gg 1$) as well as the 
continuum limit (the algorithm being implemented on a lattice). 
All the fits given below, as well as data points, are given with a precision better than $\pm0.02E_F$ (for $k_Fl\geq1.5$) and $\pm0.04E_F$ (for $0.3<k_Fl<1.5$).  The data have been averaged on $50$ to $200$ 
samples for the strongest disorders.

\begin{figure}
\psfrag{AAA}{\large $1/\sqrt{k_Fl}$}
\includegraphics[keepaspectratio,width=8.5cm]{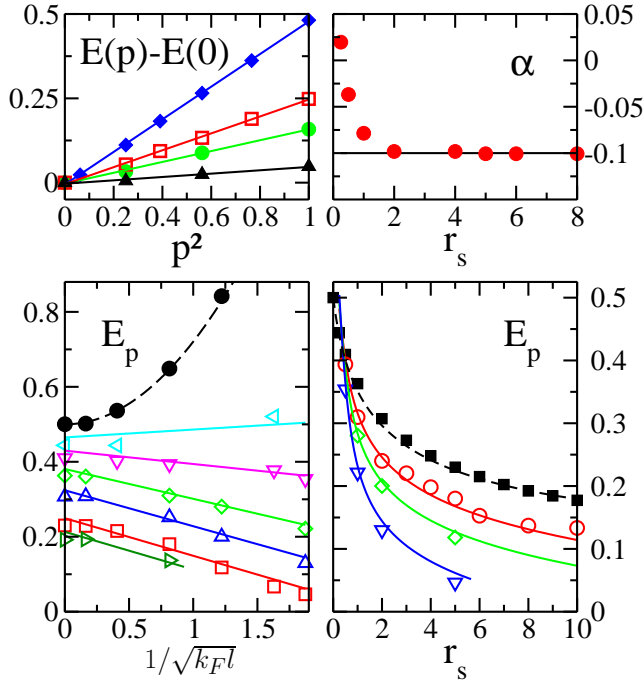}
\caption{\label{Ep} 
Polarization energy.
Upper left panel: $E(p)$ as a function of $p^2$, for $1/k_Fl=0$, $r_s=0$ (diamonds), $1/k_Fl=0$, $r_s=10$ (circles), $1/k_Fl=0.66$, $r_s=2$ (squares) and $1/k_Fl=2.65$, $r_s=6$ (triangles). Straight lines are linear fits. 
Upper right panel: $\alpha$ as defined in Eq.(\ref{eq:Ep2}) as a function of $r_s$.
Lower left panel: $E_p$ as a function of disorder $1/\sqrt{k_Fl}$, for $r_s=0$ (circles), $r_s=0.25$ (left triangles), $r_s=0.5$ (down triangles), $r_s=1$ (diamonds), $r_s=2$ (up triangles), $r_s=5$ (squares) and $r_s=8$ (right triangles). Dashed line stands for $E_p/E_F=1/2+\log2/(\pi k_Fl)$. 
Lower right panel: $E_p$ as a function of $r_s$ for $1/k_Fl=0$ (squares), $1/k_Fl=0.66$ (circles), $1/k_Fl=1.49$ (diamonds) and $1/k_Fl=3.51$ (triangles). Dashed line is the result of Ref.~\cite{marchi2008} 
without disorder. In both lower panels, the lines are fits given by Eq.(\ref{eq:Ep2}) for $r_s\geq0.25$.
All energies are in unit of $E_F$.
}
\vskip -0.4cm
\end{figure}

At very high density, i.e. no disorder ($1/k_Fl=0$) and interaction ($r_s=0$), it is straightforward 
to show that the energy depends on the polarization $p$ in a quadratic manner: 
$E(p)=E_F(1+p^2)/2$. In fact, we find that such a quadratic dependence is verified with good
precision for a wide range of parameters $0\leq r_s\leq10$ and $k_Fl>1$, corresponding to the disorder
regime of the experiments, as shown in  the upper left panel of Fig.~\ref{Ep}. 
We found (small) deviations from the quadratic 
behavior for rather large disorder in the localized regime (see triangles  
at $1/k_Fl=2.65$ and $r_s=6$). Hence we find with good precision that, 
\be
\label{eq:E_de_p}
E(p)=E(0)+E_p p^2~,
\ee
where the coefficient $E_p(r_s,k_Fl)$ is the polarization energy of the system and is proportional to the Fermi energy. 
Without electron-electron interaction and disorder $E_p=E_F/2$, and as interaction is turned on, $E_p$ decreases (indeed a ferromagnetic state tends to minimize Coulomb energy as antisymmetric orbital wave-functions have a lower probability for two electrons to be close to each other than symmetric ones).

In presence of an {\it in-plane} (i.e. purely Zeeman) magnetic field $B$, Eq.(\ref{eq:E_de_p}) turns into
$E(p)=E(0)+E_p p^2 - g \mu_B B p/2 $ ($g=2$ Lande factor and $\mu_B$ Bohr magneton). Introducing $B_p=4E_p/(g\mu_B)$,
the minimum of energy is found for a polarization $p^*$ given by,
\begin{eqnarray}
p^*&=& B/B_p \ \ \ \ B<B_p,\\
p^*&=&1 \ \  \ \ \ \ \ \ \ \   B>B_p.
\end{eqnarray}
Hence, the spin susceptibility $\chi=n_s/2 (dp/dB)_{B=0}$ is directly related to the field $B_p$ at which the polarization saturates. Defining the non-interacting susceptibility $\chi_0=g\mu_B m^*/(2\pi\hbar^2)$ , 
one finds $E_F/E_p=2\chi/\chi_0$. In addition to the characteristic magnetic field $B_p$, $E_p$ is also associated with the characteristic temperature $kT_p = E_p$ at which the polarized excited states get significantly populated. 

In the lower panels of Fig.~\ref{Ep}, we plotted our $E_p$ data as a function of disorder 
$1/\sqrt{k_Fl}$ (left) and interaction $r_s$ (right). 
Without disorder (squares in the lower right panel), our results are in 
close agreement with previous Diffusion Monte Carlo calculations performed in Ref.~\cite{marchi2008} (dashed line). As expected, the polarization energy $E_p$ decreases with interaction $r_s$. However, contrary to the single-valley case where $E_p$ becomes extremely small and could even become negative (ferromagnetic instability), 
we find that $E_p$ is always positive up to $r_s=20$ where $E_p=0.13 E_F$.\\

We find that disorder tends to increase $E_p$ for very small interactions  $r_s\leq0.25$. At $r_s=0$,
the disorder correction to $E_p$ is very well described by the second order perturbative correction
$E_p/E_F=1/2+\log2/(\pi k_Fl)$ (dashed line in lower left panel of Fig.~\ref{Ep}). However, as soon as small interactions are switched on ($r_s>0.3$),
disorder tends to decrease $E_p$. We also find that the second order perturbative correction 
$\propto 1/k_Fl$ is valid only for very tiny disorder $1/k_Fl<0.04$ above which the correction
becomes proportional to $1/\sqrt{k_Fl}$. In order to make our set of data easily retrievable, we use the following fit, valid for
 our entire
data set for $0.25<r_s<10$,
\be
\label{eq:Ep2}
E_p(k_Fl,r_s)=E_p^{\rm cl}(r_s)+\frac{\alpha(r_s)}{\sqrt{k_Fl}}\,E_F+ A_5 E_F,
\ee
where $E_p^{\rm cl}$ is the polarization energy of the clean system fitted (for $0<r_s<20$) with the Pade formula:
\be
\label{eq:Ep}
E_p^{\rm cl}(r_s)=\frac{A_0+A_1\,r_s}{A_2+A_3\,r_s+A_4\,r_s^2}\, E_F~.
\ee
The values of fitting parameters $A_i$ are listed in Table~\ref{tab}. 
The parameter $\alpha$ is plotted in the upper right part of Fig.~\ref{Ep} and is equal with good
precision for $r_s\ge 2$ to $\alpha=-0.1$. The small parameter $A_5$ roughly accounts for the fact that at very low disorder, the
disorder-correction to $E_p/E_F$ is not linear but quadratic in $1/\sqrt{k_Fl}$. We note that Eq.(\ref{eq:Ep2}) and (\ref{eq:Ep}) are merely a compact way of describing our numerical data.

\begin{table}
\begin{center}
\begin{tabular}{|*{7}{p{0.8cm}|}}
\hline
 i    & $0$ & $1$ & $2$ & $3$ & $4$ & $5$\\
\hline
$A_i$ & $27.93$ & $9.83$ & $56.5$ & $46$ & $1.77$ & 0.019 \\
\hline
\end{tabular}
\end{center}
\vspace{-12pt}
\caption{\label{param_fit}
Parameters $A_i$ of Eq.~(\ref{eq:Ep2}) and (\ref{eq:Ep}).}
\label{tab}
\vskip -0.4cm
\end{table}

\section{Comparison of the polarization energy $E_p$ with the experimental characteristic energy scale.}\label{exp2}
Fig.~\ref{rho_exp} shows raw experimental data $\rho(B)$ and $\rho(T)$ on low density 
Si-MOSFETs from six different experiments,
including the original set of data from Kravchenko et al. For each sample, we estimate the parameter
$\eta$ from the peak mobility $\mu^{\rm peak}$ obtained at high density (hence probably overestimating $\eta$)
 which sets how $\sqrt{k_Fl}=\eta/r_s$
evolves with density. Then, for each density, we calculate the polarization energy $E_p$
according to Eq.~(\ref{eq:Ep2}) and (\ref{eq:Ep}). Disorder corrections to $E_p$ are rather small for those
high mobility samples, so that our results are dominated by $E_p^{\rm cl}(r_s)$ and 
not very sensitive to our estimate of $\eta$. 
For each curve $\rho(B)$ ($\rho(T)$), we draw a circle at the calculated saturation field $B_p$ (characteristic 
temperature $T_p$). The calculations are performed without any adjustable parameters and we find an
extremely good match between $B_p$ ($T_p$) and the field (temperature) at which the resistivity saturates 
(has its inflexion point). 

\subsection{Magneto-resistance experiments}
Describing the full $\rho(B)$-dependence is a complicated task. However, as the only effect of an {\it in-plane} magnetic field is to polarize the sample through Zeemann coupling (the orbital effect of an {\it in-plane} field can be neglected
with good approximation), one expects that when the polarization saturates, the resistivity also saturates~\cite{okamoto1999}, so that
 $\rho(B)$ provides a direct measurement of $B_p$.

The experimental data of resistivity as a function of in-plane magnetic field $B$ (at low temperature) shown in panels (d), (e) and (f) of Fig.~\ref{rho_exp} correspond to three samples with mobility $\mu^{peak}=25000\, \rm{cm^2/Vs}$ \cite{mertes1999}, $41000 \,\rm{cm^2/Vs}$ \cite{pudalov1997} and $20000 \,\rm{cm^2/Vs}$ \cite{vitkalov2001}. We find that the magnetic field $B_{sat}$ at which the resistivity saturates is in extremely good agreement with our calculated $B_p$. This agreement between the experimental $B_{sat}$ and the calculated $B_p$ can be considered as a 
validation of the minimum model that we have used as well as of the accuracy of the QMC method.

\begin{figure}
\includegraphics[keepaspectratio,width=8.5cm]{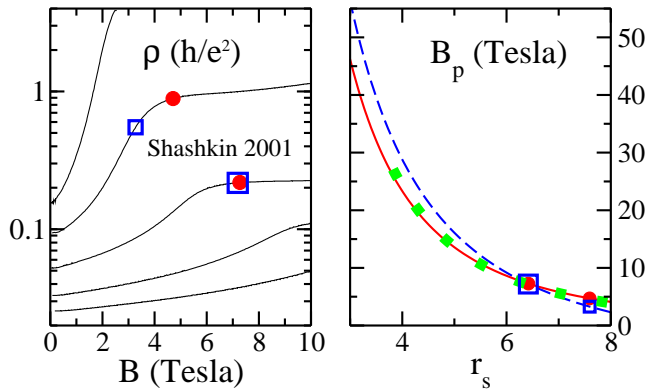}
\caption{\label{Bc} 
Left panel: $\rho$ as a function of $B$ at different $n_s$, from Ref.~\cite{shashkin2001}. Right panel: experimental $B_{sat}$ from Ref.~\cite{shashkin2001} (dashed line), QMC $B_p$ (full line) and experimental $B_p$ from Ref.~\cite{okamoto1999} (green squares) as a function of $r_s$. In both panels, red points correspond to QMC $B_p$ and blue squares to $B_{sat}$ from Ref.~\cite{shashkin2001}, at two fixed densities.}
\vskip -0.4cm
\end{figure}

Let us now discuss a publication~\cite{shashkin2001} which claimed to observe a divergence of the spin susceptibility $\chi$. 
Such a finding would contradict our calculation as $\chi\propto 1/B_p$, and although $B_p$ decreases as one lowers the density, we did not observe any divergence in the numerical calculations (one should however keep in mind that in the lower right panel of Fig.\ref{Ep}, the energies are in unit of $E_F$ which itself is proportional to the electronic density $n_s$).
The corresponding data are shown in the right panel of Fig.~\ref{Bc} (our $B_p$: full line, experimental $B_{sat}$: dashed line). The experimental $B_{sat}$ were obtained in Ref.~\cite{shashkin2001} from rescaling the in-plane magneto-resistivity data ($\mu^{peak}=30000 \,\rm{cm^2/Vs}$) at small field. Here, we find a good, but not very good, agreement between $B_p$ and $B_{sat}$. To understand the discrepancy, let us come back to the raw $\rho(B)$ experimental data for two values of the density corresponding to the blue squares (experiments) and red circles (our numerics). Those raw data are shown on the left panel of Fig.~\ref{Bc} where we also reproduce our predictions as well as the value of $B_{sat}$ obtained from the rescaling procedure. Clearly, the rescaling procedure underestimates $B_{sat}$ as the resistivity has not saturated yet at this value while our calculated $B_p$ matches precisely the end of the resistivity rise. We find that the agreement between the experiments and the numerics is actually extremely good and that the rescaling procedure only give approximate estimates of $B_{sat}$. As the
divergence of $\chi$ claimed in Ref.\cite{shashkin2001} is extremely sensitive to this procedure, we conclude that there is probably no divergence of the susceptibility as defended in Refs.~\cite{pudalov2001,prus2003} but in contradiction with Ref.~\cite{shashkin2001}. New thermodynamic measurements are in qualitative\cite{reznikov2009} and quantitative\cite{shashkin2006} agreement with our numerics (hence we disagree with the conclusions of Ref.\cite{shashkin2006} which claim that there is a divergence of the spin susceptibility).

Finally, we add a last data set that corresponds to the measures made in Ref.~\cite{okamoto1999} (green squares in right panel
of Fig.~\ref{Bc}) from the analysis of Shubnikov de Haas oscillations, at a similar mobility ($\mu^{peak}=24000 \,\rm{cm^2/Vs}$). We find that they agree perfectly with our calculations. Overall, we have a quantitative agreement between $B_p$ and $B_{sat}$ for five different experiments indicating that our minimum model and QMC calculations adequately describe the experimental situation.

\subsection{Resistance versus temperature experiments}
We now turn to the original puzzle, the resistivity versus temperature $\rho(T)$ experiments. A first remark that was originally 
made in Ref.~\cite{altshuler2000bis} is that the large increase of resistance takes place at large temperatures compare to usual transport experiments. This can be seen for instance in Fig.~\ref{rho_exp}\,(c) where temperature has been rescaled with respect to the Fermi temperature $T_F$. One finds that the inflexion point of $\rho(T)$ takes place for $T/T_F=0.2$.
Refs.~\cite{altshuler2000bis,brunthaler2001} attributed this large energy scale to some semi-classical effect. Here however, we argue that this characteristic scale is the polarization temperature $T_p$. On a qualitative level, it is not surprising: this energy scale is equal to a fraction of the Fermi temperature ($0.5 \,T_F$ without interaction and disorder and less in their presence) and we have already showed that it controls the magneto-resistance behavior.

More quantitatively, in panels (a), (b) and (c) of Fig.~\ref{rho_exp}, we plot various experimental data of resistivity $\rho$ as a function of temperature $T$, at zero magnetic field and at various electronic densities, for three Si-MOSFETs with respectively $\mu^{peak}=30000\,\rm{cm^2/Vs}$ (Ref.~\cite{kravchenko1995}), $41000 \,\rm{cm^2/Vs}$ (Ref.~\cite{pudalov1998}) and $19600 \,\rm{cm^2/Vs}$ (Ref.~\cite{prinz2000}). We have superimposed the circles corresponding to Quantum Monte Carlo polarization temperatures $T_p$. The typical temperature where one observes
the change of resistivity associated with the metallic behavior matches our calculated $T_p$. 
This is a strong indication that the polarization energy is the unique energy scale that controls the physics of the metallic behavior.

\section{A minimum scenario for the 2D Metal-Insulator transition}\label{scenario}
We now have a good understanding of what happens on the horizontal of the $\rho(T)$ and $\rho(B)$ curves. To gain a full understanding of 
properties {\bf P1} and {\bf P2}, we need to understand what happens on the vertical axis, i.e. why does the resistivity increase and not decrease upon increasing temperature or magnetic field. In other words, in order to account for 
both {\bf P1} and {\bf P2}, it is necessary and sufficient to show that {\it polarized states have a higher resistivity than 
non polarized one} (hereafter referred as property {\bf P6}).  Indeed, it is necessary, since the  $\rho(B)$ curves show that the polarized system (high field) has a higher resistivity than the non polarized one (zero field). It is also sufficient as, once {\bf P6} is established, it naturally follows that when one will increase temperature in a range around $T_p$, the highly resistive excited states will be significantly populated and the overall resistivity will increase.

\begin{figure}
\psfrag{BBB}{\large $1/\sqrt{k_Fl}$}
\includegraphics[keepaspectratio,width=8.5cm]{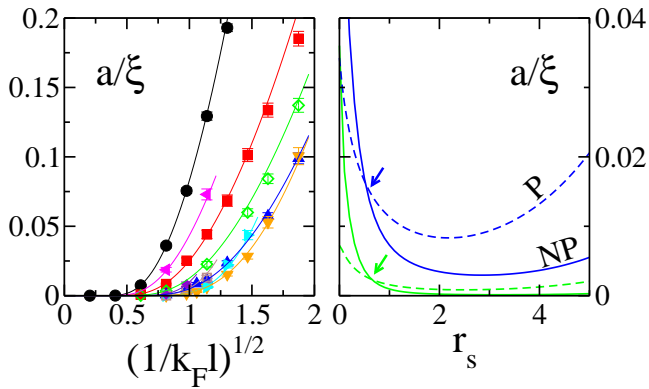}
\caption{\label{fig:xsi} 
Left panel: $a/\xi$ as a function of $1/\sqrt{k_Fl}$ for the non-polarized Si-MOSFET. The symbols correspond to different strengths of interaction: from left to right, $r_s=0$ (circles), $r_s=0.1$ (left triangles), $r_s=0.25$ (full squares), $r_s=0.5$ (empty diamonds), $r_s=1$ (up triangles), then from right to left, $r_s=2$ (down triangles), $r_s=4$ (right triangles), $r_s=6$ (empty squares) and $r_s=10$ (full diamonds). Right panel: $a/\xi$ as a function of $r_s$ for the non-polarized (NP, full lines) and polarized (P, dashed lines) Si-MOSFET, at $k_Fl=1.51$ (two lower green curves) and $k_Fl=0.89$ (two upper blue curves). The arrows indicate the interaction strength $r_s$ at which the polarized excited state becomes more localized than the non polarized ground state. In both panels, the lines are given by Eq.~(\ref{eq:fit_xi0}), (\ref{eq:fit_xi1}) and (\ref{eq:fit_xi2}).
}
\vskip -0.4cm
\end{figure}

In a previous publication\cite{fleury2008bis}, we have performed a systematic study of the interplay between Anderson localization and electron-electron interactions in Si-MOSFETs. We found that upon increasing interactions, the localization length of the non polarized ground state strongly 
increases (in absolute value, but in particular with respect to the polarized excited states), so that one naturally accounts for {\bf P6}. 
This result is also shown in the right panel of Fig.~\ref{fig:xsi} where we compare as a function of $r_s$ the localization lengths $\xi$ (in unit of the average distance between electrons $a$) computed at zero temperature for the non-polarized electron gas and for the fully polarized one. At very weak interaction ($r_s\lesssim 0.5$), the polarized system is less localized than the non-polarized one.
Indeed, polarizing the system raises the Fermi level $E_F$ (by a factor $2$), hence also $k_Fl$, hence $\xi$ which depends exponentially on the latter. On the contrary, at stronger interaction  ($r_s\gtrsim 0.5$), the situation is reversed and the polarized system is more localized than the non-polarized one. We checked that we also get such an inversion of the localization lengths between the non-polarized ground state and \textit{partially} polarized states, so that we indeed recover {\bf P6}. In Fig.4 of Ref.\cite{fleury2008bis} we presented the effective phase diagram constructed out of this mechanism which should be compared with the experimental one presented in Fig.~\ref{fig_diagphaseexp}. The agreement is semi-quantitative,  meaning that our scenario also captures properties {\bf P3} and {\bf P4} (as well as {\bf P5} discussed above). Hence, we find that the minimum model discussed in this paper is enough to account for all the relevant experimental facts discussed in the beginning of this paper.

We end this section with a discussion of the existence of "true" metal at zero temperature, i.e. does the localization length diverge in the region where the metallic behavior is observed. 
We emphasize that in light of the above discussion, this is a rather academic question as what has  actually been measured 
experimentally corresponds to rather high energy physics.
In Ref.\cite{fleury2008bis}, we have claimed that there probably does not exist a "true" metal at zero temperature. Our main argument was the absence of any visible deviation to the one-parameter scaling theory\cite{abrahams1979} in presence of interactions, despite a huge delocalization effect. Here we give another argument. In the left panel of Fig.~\ref{fig:xsi}, we plot $a/\xi$ as a function
of the disorder strength $1/\sqrt{k_Fl}$ for various values of $r_s$. Without interaction\cite{lee1985}, the localization length of the two-dimensional electron gas depends exponentially on $k_Fl$: $\xi/a=(k_Fl/B)\exp(Ck_Fl)$. We find that in the presence of interactions our entire set of data can be fitted by the same law upon renormalizing the $B$ and $C$ parameters. Hence, while the localization length can become large in the presence of interaction, it always remains finite and the system remains ultimately an insulator. In practice this insulating behavior should appear at extremely low temperature when the phase coherence length becomes larger than the localization length. Such a metal-to-insulator re-entrance as a function of temperature has been observed at high density\cite{hamilton1999,pudalov1999} and  at low density, in the vicinity of the so-called metal-insulator transition, in a GaAs two-dimensional hole gas\cite{huang2008} and in Si-MOSFETs\cite{prus2001}.

Below we provide the fitting parameters that account for our localization length data. We have
\be
\label{eq:fit_xi0}
\frac{a}{\xi}(k_Fl,r_s)=\frac{B(r_s)}{k_Fl}\exp\left(-C(r_s)k_Fl\right)\,,
\ee
where two different sets of parameters are needed for the non-polarized  $(B^{NP},C^{NP})$ and the polarized $(B^{P},C^{P})$ case.
In turn, the parameters $B$ and $C$ are well fitted by:
\begin{eqnarray}
\label{eq:fit_xi1}
B(r_s)&=&b_0+b_1 \sqrt{r_s}+b_2\, r_s + b_3\, r_s \sqrt{r_s}\nonumber \\ 
      & &{}+ b_4\, r_s^2 + b_5\, r_s^2 \sqrt{r_s} +b_6\, r_s^3
\end{eqnarray}
and:
\begin{eqnarray}
\label{eq:fit_xi2}
C(r_s)&=&c_0+c_1 \sqrt{r_s}+c_2\, r_s + c_3\, r_s \sqrt{r_s}\nonumber \\ 
      & &{}+ c_4\, r_s^2 + c_5\, r_s^2 \sqrt{r_s}+c_6\, r_s^3\,,
\end{eqnarray}
where the values of the $b_i$ and $c_i$ fitting parameters are given in Table~\ref{param_fit_xsi} for the non-polarized system ($b_i^{NP}$ and $c_i^{NP}$) and for the polarized one ($b_i^{P}$ and $c_i^{P}$), with the convention that $k_Fl$ always refers to the Fermi level of the non-polarized gas. Eq.~(\ref{eq:fit_xi0}) supports the idea that the two-dimensional electron gas behaves as a Fermi liquid (at least for moderate interactions $r_s\leq 10$) and that the effect of electron-electron interactions is to renormalize its effective characteristics.

\begin{table*}
\begin{center}
\renewcommand{\arraystretch}{1.2}    
\begin{tabular}{|c|c|c|c|c|c|c|c|}
\hline
 i    & $0$ & $1$ & $2$ & $3$ & $4$ & $5$ & $6$\\
\hline
$b_i^{NP}$ & $0.191$ & $-0.325$ & $0.24$ & $-0.0526$ & $0.0035$ & $0$ & $0$\\
\hline
$b_i^{P}$ & $0.136$ & $-0.09093$ & $-0.03197$ & $0.18232$ & $-0.17749$ & $0.07549$ & $-0.010827$\\
\hline
$c_i^{NP}$ & $0.84$ & $-0.177$ & $1.094$ & $2.0616$ & $-1.6784$ & $0.3027$ & $0$\\
\hline
$c_i^{P}$ & $1.68$ & $-0.712$ & $2.418$ & $-1.1874$ & $0.1594$ & $0$ & $0$\\
\hline
\end{tabular}
\end{center}
\vspace{-12pt}
\caption{\label{param_fit_xsi}
Parameters $b_i$ and $c_i$ of Eq.~(\ref{eq:fit_xi1}) and (\ref{eq:fit_xi2}) for the non-polarized Si-MOSFET ($b_i^{NP}$ and $c_i^{NP}$) and for the polarized one ($b_i^{P}$ and $c_i^{P}$).}
\label{tab}
\vskip -0.4cm
\end{table*}

\section {Conclusion}\label{conclusion}

Let us discuss a few theoretical scenarios that have been proposed by other authors. In light of the findings of this paper, we believe that models whose characteristic energies do not involve the polarization energy should not be applied to those experiments. This includes for example the percolation scenario proposed in Ref.\cite{meir1999} (although percolation probably plays an important role close to the insulating region in GaAs heterostructures). 
This also includes the proposal related to Wigner crystal Ref.\cite{spivak2003} (which we dismissed in Ref.~\cite{waintal2006} on other grounds). Ref.\cite{camjayi2008} has some overlap with our scenario
(on the role of polarization energy in particular) but relies on the existence of some form of local order (which we did not observe in our simulations) and does not include the presence of disorder. Extrinsic models as the temperature dependent disorder scenario proposed in Refs.\cite{altshuler1999,altshuler2000ter} can also be excluded on the same ground (but these authors were the first to recognize the presence of a high energy scale in the experiments). An important theoretical advance was made in Ref.\cite{punnoose2005} where the authors studied the interplay between interaction and disorder in the diffusive limit, in the limit of an infinitely large number of valleys. We believe that those authors correctly pointed out the important role of valley degeneracy in the experiments, at the origin of the large differences between the behaviors observed in Silicon MOSFETs and in other heterostructures.
However, the limit of an infinite number of valleys artificially increases the role of electron-electron interactions. In particular those authors find that even a very weak interaction will drive the system toward a non Fermi-liquid fixed point. Those results, which imply a complete breakdown of one parameter scaling theory are in contradiction with our numerical results\cite{fleury2008bis}. More importantly, the corresponding analysis of the experiments\cite{anissimova2007,knyazev2008} is in contradiction with the present analysis of the characteristic energy scales involved in these systems.

In summary, we have considered a {\it minimum } model of Si-MOSFETs taking into account
Coulomb repulsion, spin, valley degeneracy, and disorder in a non perturbative way. Our chief result
is that the polarization energy $E_p$ calculated for this model is in quantitative agreement with the
characteristic energy scale that controls the metallic behavior of those high mobility Si-MOSFETs.
Beside spin, Coulomb repulsion plays a crucial role here as it decreases $E_p$ by a factor $2-3$ (see 
Fig.\ref{Ep}). The presence of valley degeneracy is also very important as it also decreases $E_p$ by a factor $2$. On the other hand, disorder only gives corrections of $10-30\%$ to $E_p$ for the samples considered in this survey. We note that the polarization temperature is the ``crossover temperature'' that was forseen in Ref.~\cite{altshuler2000bis}. However, we find that this ``crossover temperature'' is an intrinsic property of the electron gas, and does not involve any additional extrinsic ingredient. 
A simple corollary of our chief result is that the non-polarized system has to be a much better conductor 
than the polarized one (to explain the sign of $d\rho/dT$ and $d\rho/dB$ in the metallic region).  We have argued 
before\cite{fleury2008bis} that our minimum model accounts for this counter-intuitive point ({\bf P6}), as the interplay between Coulomb repulsion and Anderson localization depends strongly on the polarization. These findings imply that our ``necessary'' model (in the sense that all its ingredient are known to be present and {\it a priori} relevant) is also ``sufficient'' to capture the essential features of the metallic behaviors. 
Another consequence of our model is the apparent failure of one parameter scaling. {\it For a given polarization}, we could not find any deviation to the (one parameter) scaling theory of localization\cite{fleury2008bis}. However, at finite temperature excited states of different localization lengths come into play so that the physics is no longer controlled by a single parameter.
In this work, we have focused on Si-MOSFETs, but at the qualitative level,  many statements also apply to other materials like GaAs 
heterostructures. In our view, the situation in the latter is made a bit different by the conjunction of three elements. First the absence of valley degeneracy makes the delocalization effect of the ground state much less effective. Second, the present of the doping layer close to the gas induces some long range disorder which can mask the local physics. Last, extremely high mobility are available for these systems, making it possible to study almost ballistic samples.

\acknowledgments
We thank S. De Palo and G. Senatore for interesting discussions as well as M. Reznikov for valuable comments on the manuscript. Support from the CCRT supercomputing facilities is also acknowledged.

\bibliography{energy}

\end{document}